\RequirePackage{amsmath}
 \documentclass[12pt]{iopart}
\usepackage{bm,bbm}
\usepackage{mathrsfs}
\usepackage{graphicx}
\usepackage{color}
\usepackage{amsmath}
\usepackage{amssymb,amsfonts}
\usepackage{mathtools}
\usepackage{tikz}
 \usetikzlibrary{calc,arrows,decorations.pathreplacing,backgrounds}
\colorlet{mylinkcolor}{blue!66!black!80}
\usepackage[colorlinks=true,linkcolor=mylinkcolor,citecolor=mylinkcolor,filecolor=cyan,urlcolor=mylinkcolor,breaklinks=true]{hyperref}
\usepackage{xcolor}

\usepackage[utf8]{inputenc}
\usepackage{booktabs}
\usepackage{bbm,mathrsfs,mathtools}

\begin{document}

\title[Exact Single- and Two-Tag Local Times Beyond Large Deviation
  Theory]{Unfolding tagged Particle Histories in Single-File Diffusion:
  Exact Single- and Two-Tag Local Times Beyond Large Deviation Theory}

\author{Alessio Lapolla$^{\dagger}$ and Alja\v{z} Godec$^{\dagger}$}
\address{$\dagger$ Mathematical Biophysics Group, Max-Planck-Institute for
  Biophysical Chemistry, G\"{o}ttingen 37077, Germany}
\ead{agodec@mpibpc.mpg.de}

\begin{abstract}
  Strong positional correlations between particles render the
  diffusion of a tracer particle in a single file anomalous and
  non-Markovian. While ensemble average
  observables of tracer particles are nowadays well understood, little
  is known about the statistics of the corresponding functionals, i.e.
  the time-average observables. It even remains unclear
  how the non-Markovian nature emerges from correlations between
  particle trajectories at different times.
  Here, we first present rigorous results for fluctuations and two-tag correlations of
general bounded functionals of ergodic Markov processes with a
diagonalizable propagator. They relate the statistics
of functionals on arbitrary time-scales to the relaxation
eigenspectrum. Then we study tagged particle local times -- the time a
tracer particle spends at some predefined location along a single trajectory up to a time $t$. 
Exact results are derived for one- and two-tag local
times, which reveal how the individual particles' histories become
correlated at higher densities because 
each consecutive displacement along a trajectory requires collective
rearrangements.  
Our results unveil the intricate
meaning of projection-induced memory
on a trajectory level, invisible to ensemble-average
observables, and allow for a detailed analysis of single-file experiments
probing tagged particle exploration statistics.
\end{abstract}


\section{Introduction}
Single-file dynamics refers to the motion of particles in a narrow,
effectively one-dimensional channel, which prevents their
crossing, and is central to the transport in
biological channels \cite{Biological} the kinetics of transcription
regulation \cite{sliding}, transport in zeolites \cite{zeolites} and
in superionic conductors \cite{superionic}. Recent advances in
single-particle tracking and nanofluidics enabled experimental
studies of single file dynamics in colloidal systems, which directly probe
the fundamental physical principles of tagged particle motion to
an unprecedented precision \cite{traj,soft}.
\begin{figure}[ht]
  \begin{center}
\includegraphics[width=14.cm]{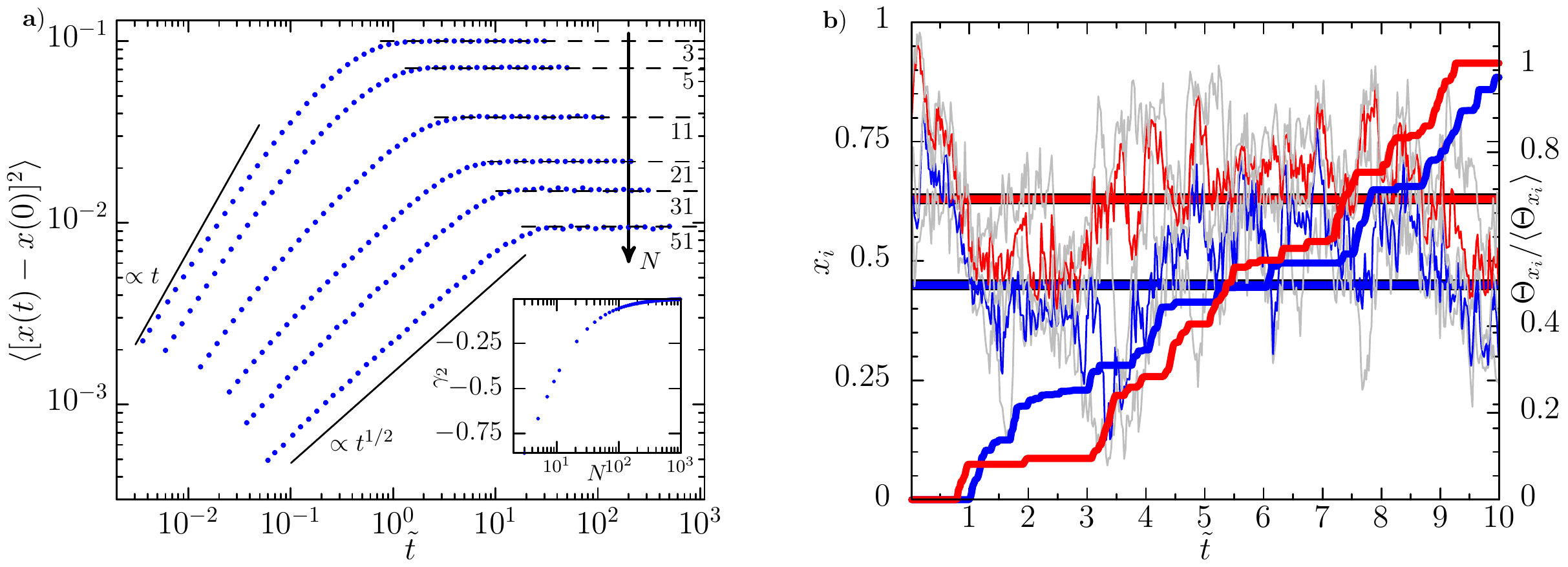}
\caption{a) MSD of the central particle in a single file with
increasing particle number $N$ starting from equilibrium initial conditions. Time is
measured in units of the mean number of collisions
$\tilde{t}=Dt/N^2$. Inset: Kurtosis excess of the invariant
measure of the central particle depending on $N$; b) Trajectories of two next-nearest
neighbor particles in a single file of $11$ particles (red and blue curves) alongside
the respective left and right nearest neighbors (gray curves). Overlaid are corresponding local time fractions up to a time $t$,
$\theta_t^i$ in the respective red and blue shaded intervals. The
remaining particle trajectories are omitted for convenience.}
\label{fg1}
 \end{center}
\end{figure}

The motion of particles in a single file is strongly correlated, which gives rise to a rich and intricate
phenomenology. In a Brownian single file the non-crossing constraint leads
to subdiffusion with the ensemble
mean squared displacement (MSD) of a tagged particle
scaling as
$\langle[x(t)-x(0)]^2\rangle\propto \sqrt{t}$ \cite{Harris}.
When confined to a finite interval
the subdiffusive scaling of the MSD is transient, saturating at an equilibrium variance, with the extent of the subdiffusive regime growing with
the particle density (see Fig.~\ref{fg1}a and
\cite{Tobias_PRLE}). Concurrently, an effective harmonization emerges at increasing density,
with the invariant measure of a tagged particle approaching a Gaussian
and
a vanishing kurtosis excess $\gamma_2=\langle x^4\rangle_{\mathrm{eq}}/\langle
x^2\rangle_{\mathrm{eq}}^2-3$ (see inset of Fig.~\ref{fg1}a).
More generally it holds that the MSD of a tagged particle in an unconfined single file and the absolute
dispersion of a free particle in the limit $t\to\infty$ are related via $\langle
[x(t)-x(0)]^2\rangle\propto \langle
|x(t)|\rangle_{\mathrm{free}}$ \cite{Percus}.
The motion of particles
on a many-body level is Markovian, the resulting tagged particle
dynamics is, however, highly non-Markovian \cite{Tobias_PRLE}, and displays a staggering dependence
on the respective initial conditions \cite{initial}.

Tremendous effort has been made to study the tagged particle dynamics
theoretically \cite{theory}. In particular, the tagged particle ensemble
propagator has been studied using the 'reflection principle'
\cite{reflection}, Jepsen mapping \cite{Eli}, momentum Bethe
ansatz \cite{Tobias_PRLE}, harmonization techniques \cite{harmon}, and
macroscopic fluctuation \cite{MFT}
and large deviation \cite{LDEV} theory. Notwithstanding, these works, with isolated exceptions \cite{Olivier},
focused on ensemble-average properties alone. State-of-the-art
experiments, however, albeit probing particle trajectories
and thereby providing direct access to functionals of paths,
are typically analyzed using ensemble-average concepts (see e.g. \cite{traj,soft}).
The analysis of functionals of tagged particle trajectories is thus not only feasible but also more natural than
studying ensemble-average observables. Moreover, to arrive at a
deeper physical understanding of projection-induced memory effects and resulting
non-Markovianity, an understanding of the correlations of particle
histories and their decorrelation on ergodic time-scales is required.    

In particular, we here focus on
the trajectory-, or time-average analogue of the tagged particle
ensemble propagator \cite{Tobias_PRLE}. Any time-average observable
can be constructed from the local time fraction (see Eq.~(\ref{auxx})
in Appendix A), which is defined as (see Fig.~\ref{fg1}b)
\begin{equation}
\theta^j_t(y)=t^{-1}\int_0^t\mathbbm{1}^j_y[\mathbf{x}(\tau)]d\tau,
\label{local}
\end{equation}
where $\mathbbm{1}^j_y[\mathbf{x}(\tau)]=1$ if
$x_j\in dy$ centered at $y$, and zero otherwise
\cite{Varadhan}. $\theta^j_t(y)$ in Eq.~(\ref{local})
is a random quantity denoting the fraction of the local time,
$t\theta^j_t(y)$ -- the time the tagged particle $j$ spends in
an infinitesimal region around the point $y$ along a trajectory up
until time $t$. $\mathbf{x}(t)\equiv(x_1(t),\ldots,x_N(t))^T$ denotes the many-body
trajectory written in vector form. The dynamics of a tagged particle
$x_{i}(t)$  irrespective of the other $N-1$ is not Markovian, and any
two tagged particle trajectories
$x_{i}(t)$ and $x_{j}(t)$ are correlated on all but ergodically long
times.  We focus on the fluctuations and two-tag correlations of local
time fractions 
\begin{eqnarray}
  \label{var}
  \sigma^2_{x_i}(t)&=&\langle \theta^i_t(x)^2\rangle-\langle
  \theta^i_t(x)\rangle^2\\
  \label{cor}
  \mathcal{C}^{ij}_{xy}(t)&=&\langle \theta^i_t(x)\theta^j_t(y)\rangle-\langle\theta^i_t(x)\rangle\langle\theta^j_t(y)\rangle,
\end{eqnarray}  
where $\langle \cdots \rangle$ denotes the average over all
$N$-particle trajectories starting from the steady-state (in this case
Boltzmann equilibrium) and propagating up
to time $t$. Note that for ergodic Markov dynamics
$\lim_{t\to\infty}\sigma^2_{x_i}(t)=0$ and
$\lim_{t\to\infty}\mathcal{C}^{ij}_{xy}(t)=0$, reflecting the fact that
on ergodically long time-scales time-average observables become
deterministic and correlations between them vanish.

A general theory of local times in such correlated
non-Markovian dynamics so far remained elusive.  
And while the statistics of functionals of the form in Eq.~(\ref{local}) in one-dimensional stochastic
processes have been studied extensively in a variety of fields
\cite{Kac,functionals}, studies of tagged particle functionals in
interacting many-body systems are sparse, and mostly limited to
extreme value statistics of vicious walkers (see
e.g. \cite{vicious}). 

Here, we present rigorous results for variances and
two-tag correlations of bounded functionals\footnote{We consider
  functionals $V[\mathbf{x}(t)]$ of Markovian trajectories $\mathbf{x}(t)$,
for which $V[\mathbf{x}(t)]<\infty, \forall t$ with probability 1 (see
e.g. \cite{Kac}).} of Markovian dynamics on arbitrary time-scales, in
terms of the relaxation eigenspectrum of the corresponding
propagator. The theory also covers the case, when a higher dimensional
dynamics is projected onto a smaller subspace thereby leading to
non-Markovian dynamics on the reduced subspace, a hallmark example
thereof being tagged-particle dynamics in a single file.
The theory applies to all ergodic Markovian systems with a
diagonalizable propagator. As an example we study tagged
particle local times in a single file of Brownian point particles in a
box. Diagonalizing the many-body propagator using the coordinate Bethe ansatz, our results uncover
non-Poissonian trajectory-to-trajectory fluctuations of local times, and a cross-over from negatively to positively correlated two-tag particle
histories upon increasing density, mirroring the emergence of
collective fluctuations breaking Markovianity in tagged particle
motion and leading to tracer subdiffusion. Clear and long-lived
deviations of local time statistics from shot-noise behavior
demonstrate the insufficiency of harmonization concepts for describing
tracer diffusion on a trajectory level.   
More generally, the connection to the relaxation spectrum provides an intuitive
understanding of non-Poissonian statistics at sub-ergodic times in a
general setting.      

\section{General theory} We consider a trajectory of a general $N$-dimensional
system $\mathbf{x}(t)$ evolving according to Fokker-Planck or discrete-state
Markovian dynamics.  We are interested in ergodic
systems with a unique steady-state $\overline{P}(\mathbf{x})$ and also assume steady state
initial conditions. Due to ergodicity the mean local
time fraction $\langle \theta^j_t(y) \rangle$ under these conditions
is independent of $t$\footnote{On the level of the mean alone the
  time-ordering in the functional in Eq.~(\ref{local}) is not important
  (for a proof see Eq.~(\ref{ergmean}).} and coincides with the invariant measure $\langle \theta^j_t(y) \rangle=\int
d\mathbf{x}^{N}\delta(y-x_j)\overline{P}(\mathbf{x})$, where we
introduced the Dirac delta function $\delta(x)$ (for a proof see Eq.~(\ref{ergmean})). In the presence
of detailed balance (DB) $\overline{P}(\mathbf{x})$ is the
Boltzmann-Gibbs  measure $P_{\mathrm{eq}}(\mathbf{x})$. 

Obtaining Eqs.~(\ref{var}-\ref{cor}) essentially amounts to computing the probability generating function of the
joint local time functional given by the Feynman-Kac path integral
\begin{equation} 
  \label{joint}
Q_{u,v}(x^i,y^j|t)=\hat{\mathcal{L}}_u^{\vartheta^i}\hat{\mathcal{L}}_v^{\vartheta^j}\langle\delta(\vartheta^i-t\theta^i_t(x))\delta(\vartheta^j-t\theta^j_t(y))\rangle,
\end{equation} 
where we introduced  the
Laplace transform $\hat{\mathcal{L}}_s^{\vartheta}f(\vartheta)=\int_0^\infty
d\vartheta\mathrm{e}^{-s\vartheta}f(\vartheta)$. The moments in Eqs.~(\ref{var}-\ref{cor})
are
obtained from $\langle
\theta^i_t(x)^n\theta^j_t(y)^m\rangle=t^{-2}\partial^{n}_{v}\partial^m_{u}Q_{u,v}(x^i,y^j|t)|_{u=v=0}$
with $n+m=2$. 
A straightforward generalization of the trotterization in Ref.~\cite{Satya} shows that
$Q_{u,v}(x^i,y^j|t)$ is the propagator of a
tilted evolution operator (see Appendix A)
\begin{eqnarray}
  Q_{u,v}(x^i,y^j|t)&=&\langle \textendash|\mathrm{e}^{-t(\hat{L}+u\mathbbm{1}^i_x +v\mathbbm{1}^j_y)}|\mathrm{ss}\rangle\nonumber\\
  &=&\langle \mathrm{ss}|\mathrm{e}^{-t(\hat{L}^{\dagger}+u\mathbbm{1}^i_x +v\mathbbm{1}^j_y)}|\textendash\rangle,
\label{tilted}
\end{eqnarray}
where $\hat{L}$ and $\hat{L}^{\dagger}$ denote the 'bare' forward and
adjoint (backward) generator of the Markov process \cite{CMP}, and we introduced the 'flat' $|\textendash\rangle\equiv\int
d\mathbf{x}|\mathbf{x}\rangle$ and steady states
$|\mathrm{ss}\rangle=\int d\mathbf{x}\overline{P}(\mathbf{x})|\mathbf{x}\rangle$ in the bra-ket notation, which are the left (right) and right (left) ground eigenstates
of $\hat{L}$ ($\hat{L}^{\dagger}$), respectively.
We obtain exact expressions for the moments in
Eqs.~(\ref{var}-\ref{cor}) by performing a Dyson series-expansion of
Eq.~(\ref{tilted}) \cite{Dyson}, converging for any bounded
functional of $\mathbf{x}(t)$ (see proof in Appendix A).  

Having assumed diagonalizability of $\hat{L}^{\dagger}$ (and
$\hat{L}$) \footnote{A sufficient but not necessary condition
  guaranteeing diagonalizability is that that the operator is normal,
  i.e. commutes with its adjoint, $\hat{L}^{\dagger}\hat{L}-\hat{L}\hat{L}^{\dagger}=0$.}, we expand the backward operator in a complete bi-orthogonal set of
left and right eigenstates \footnote{Note that $\hat{L}|\psi^R_k\rangle=\lambda_k|\psi^R_k\rangle$ and $\hat{L}^{\dagger}|\psi^L_k\rangle=\lambda_k|\psi^L_k\rangle$ \cite{Gardiner}.},
$\hat{L}^{\dagger}=\sum_{k}\lambda_k|\psi ^L_k\rangle \langle\psi
^R_k|$, $\lambda_k$ denoting the (possibly degenerate) eigenvalues and $\langle\psi^L_k|\psi
^R_l\rangle=\delta_{kl}$. The details 
of the calculation of the moments are
shown in Appendix A. Obviously, $\langle
\psi^{R}_0|\mathbbm{1}^i_x|\psi^{L}_0\rangle=\overline{P}(x)$, since the
system is ergodic. The exact results for the variance and correlations
are conceptually remarkably simple and read
  \begin{eqnarray}
    \label{svar}
\!\!\!\sigma^2_{x_i}(t)\!&=&\!\!\sum_{k \ge
  1}2 \frac{\Omega_{k}(x_i,x_i)}{\lambda_k t}\left(1-\frac{1-\mathrm{e}^{-\lambda_kt}}{\lambda_kt}\right)\\
\label{scor}
\!\!\!\mathcal{C}^{ij}_{xy}(t)\!&=&\!\sum_{k \ge
  1}\!\frac{\Omega_{k}(x_i,y_j)\!+\!\Omega_{k}(y_j,x_i)}{\lambda_k t}\!\left(1\!-\!\frac{1-\mathrm{e}^{-\lambda_kt}}{\lambda_kt}\right)\!\!,
  \end{eqnarray} 
where we introduced the auxiliary function $\Omega_{k}(x_i,y_j)\equiv\langle
\psi^{R}_0|\mathbbm{1}^i_x|\psi^{L}_k\rangle\langle
\psi^{R}_k|\mathbbm{1}^j_y|\psi^{L}_0\rangle$. The exact large
deviation (LD) limits of Eqs.~(\ref{svar}-\ref{scor}) readily follow in
the limit $t\gg\lambda_1^{-1}$
  \begin{eqnarray}
    \label{LDsvar}
\sigma^{2,\mathrm{LD}}_{x_i}(t)&\simeq& 2 t^{-1} \sum_{k \ge 1}\lambda_k^{-1}\Omega_{k}(x_i,x_i)\\
\label{LDscor}
\mathcal{C}^{ij,\mathrm{LD}}_{xy}(t)&\simeq& t^{-1}\sum_{k
  \ge 1} \lambda_k^{-1}[\Omega_{k}(x_i,y_j)+\Omega_{k}(y_j,x_i)],
  \end{eqnarray}
  where $\simeq$ denotes asymptotic equality.  Analogous formulas for
  LD limits of local times not connected to a spectral expansion have also been developed (see
 e.g. \cite{SatyaDavid}). Notably, for systems obeying DB $\sigma^{2,\mathrm{LD}}_{x_i}(t)$
 sets a universal upper bound on the variance of $\theta_t$
 (compare Eqs.(\ref{svar}) and (\ref{LDsvar})).
The results
in Eqs.~(\ref{svar}-\ref{LDscor}) readily extend to arbitrary functionals $t^{-1}\int_0^t\hat{V}[\mathbf{x}(\tau)]d\tau$
with a bounded and local $\hat{V}$, by performing a
simple exchange $\mathbbm{1}^i_x\to \hat{V}$, modifying only
$\Omega_{k}(x_i,y_j)$ (see Appendix A). Eqs.~(\ref{svar}-\ref{scor}) with the aforementioned generalizations apply to all
diagonalizable $\hat{L}$, thus including all systems obeying DB, and represent our first main result. 

 Eqs.~(\ref{svar}-\ref{LDscor}) provide an intuitive understanding of local time statistics via a mapping onto relaxation eigenmodes, with
 fluctuation and correlation amplitudes proportional to the sum of transition
 amplitudes of excitations from the steady state to excited states
 and back, $\Omega_{k}(x_i,y_j)$. 
On ergodic time scales $\theta_t$ at different $t$ decorrelate, and hence
display features of shot-noise, i.e. $\sigma^2_{x_i}(t)$ and
$\mathcal{C}^{ij}_{xy}(t)$ decay inversely
proportional to the number of
independent observations of each excitation mode, $\sim \lambda_k^{-1}/t$. At finite times
$t\lesssim\lambda_k^{-1}$ shot-noise statistics are altered due to a
finite survival probability of the eigenmodes at a given $t$,
$(1-\mathrm{e}^{-\lambda_kt})/\lambda_k$ $\forall k$, setting a
hierarchy of correlation times $\lambda_k^{-1}$  (see correction
terms in brackets of Eqs.~(\ref{svar}-\ref{scor})).

\section{Local times in single-file diffusion} Consider the dynamics
of $N$ identical hard-core
interacting Brownian point particles diffusing in the
unit interval $[0,1]$, and set $D=1$ without loss of generality. The
extension to a finite
particle radius follows from a trivial change of coordinates
\cite{Tobias_PRLE}.
Let
$P(\mathbf{x}_0,t|\mathbf{x})\equiv \langle \mathbf{x}_0|\mathrm{e}^{-t\hat{L}^{\dagger}}| \mathbf{x} \rangle$
denote the $N$-particle backward propagator of the single file with
the following backward generator and $N-1$ internal non-crossing boundary
conditions:
\begin{equation}
\hat{L}^{\dagger}=-\sum_{i=1}^N\partial^2_{x_i}\quad,\quad \lim_{x_{i+1}\to
  x_i}(\partial_{x_{0,i+1}}-\partial_{x_{0,i}})P(\mathbf{x}_0,t|\mathbf{x})=0
\quad \forall i.
\end{equation}  
Confinement into a unit interval is imposed through external reflecting boundary
conditions
$\partial_{x_{1}}P(\mathbf{x_0},t|\mathbf{x})|_{x_{0,1}=0}=\partial_{x_{N}}P(\mathbf{x},t|\mathbf{x_0})|_{x_{0,N}=1}=0$.
Under
these boundary conditions we diagonalize $\hat{L^{\dagger}}$ using the
coordinate Bethe ansatz \cite{Bethe}\footnote{Note the difference with respect to the momentum-space
  Bethe ansatz solution \cite{Tobias_PRLE}, which does not diagonalize
  $\hat{L}$.} and obtain the
Bethe eigenvalues $\lambda_k=\pi^2\sum_i k_i^2$ and corresponding
left and right eigenvectors
\begin{equation}
\psi_k^L(\mathbf{x})\equiv\langle
\mathbf{x}|\psi^L_k\rangle=\sum_{\{k_i\}}\!^{'}\prod_{i=1}^N2^{(1-\delta_{k_i,0})/2}\cos(k_i\pi
x_i)
\label{states}
\end{equation}
and $\psi_k^R(\mathbf{x})\equiv\langle\psi^R_k|\mathbf{x}\rangle=m_k\psi_k^L(\mathbf{x})$ with
$\langle\psi_k^R|\psi^L_l\rangle=\delta_{k,l}$, where $m_k$ is the
multiplicity of the Bethe eigenmode $|\psi^L_k\rangle$ (see Appendix C), and $\sum_{\{k_i\}}'$ denotes the sum over all permutations
of single-particle eigenvalues with $k_i\in\mathbb{N}_0$.

The matrix elements entering $\Omega_{k}(x_i,y_j)$ follow
upon integration over the $n_l$ and $n_r$ particle
coordinates to the left and right, respectively, from the tagged
particle $i$ while strictly preserving the particle ordering
\cite{Tobias_PRLE}, yielding (see Appendix B)
$\langle \psi^{R}_k|\mathbbm{1}^i_x|\psi^{L}_0\rangle=\frac{N!}{m_k}\langle
\psi^{R}_0|\mathbbm{1}^i_x|\psi^{L}_k\rangle$ with
\begin{equation}
\langle \psi^{R}_0|\mathbbm{1}^i_x|\psi^{L}_k\rangle=\frac{m_k}{n_l!n_r!}\sum_{\{k_i\}}
\!^{'}
  {_x^1\Lambda_i^c}\prod_{j=1}^{i-1}{_x^x\Lambda_j^s}\prod_{k=i+1}^{N}{_{\phantom{1-{}}x}^{1-x}\Lambda_k^s},
\label{element}     
\end{equation}
and $\langle \psi^{R}_k|\mathbbm{1}^i_x|\psi^{L}_0\rangle=\frac{N!}{m_k}\langle
\psi^{R}_0|\mathbbm{1}^i_x|\psi^{L}_k\rangle$ . In Eq.~(\ref{element}) we have defined the auxiliary functions 
\begin{eqnarray}
{_x^y\Lambda_i^c}&=&
y\delta_{k_i,0}-(1-\delta_{k_i,0})\sqrt{2}\cos(\pi k_ix)\nonumber \\
{_x^y\Lambda_i^s}&=&
y\delta_{k_i,0}-(1-\delta_{k_i,0})\sqrt{2}\sin(\pi k_ix)/\pi
k_i.
\label{auxBethe}
\end{eqnarray}
This delivers exact results for $\sigma^2_{x_i}(t)$ and
$\mathcal{C}^{ij}_{xy}(t)$ in Eqs.~(\ref{svar}-\ref{LDsvar}). An
efficient numerical implementation of our analytical results can be
made available upon request. 
\begin{figure}[ht]
\begin{center}  
\includegraphics[width=14.3cm]{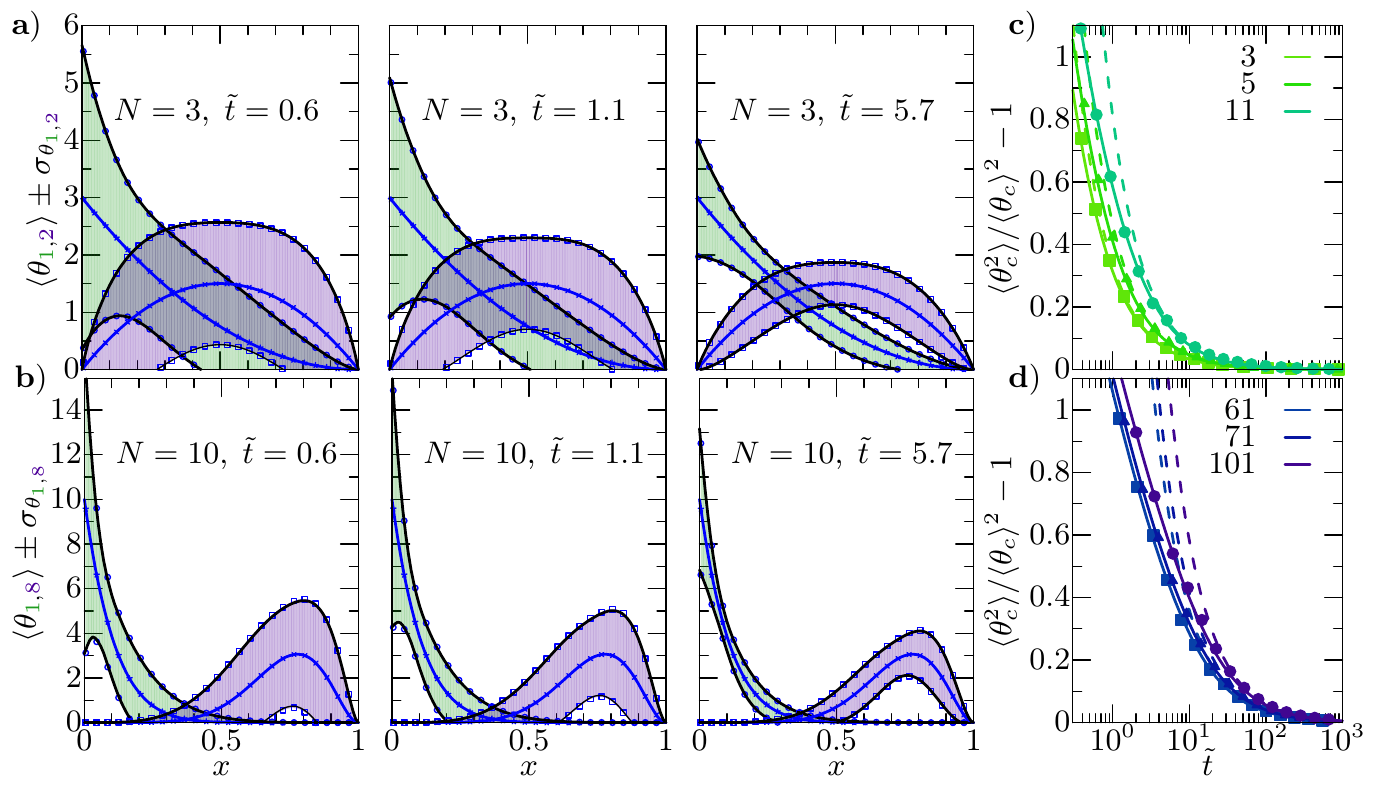}
\caption{Statistics of local time fraction: mean, $
  \langle\theta^i_t(x)\rangle$,  (blue
  lines) and fluctuations reflected by the shaded area enclosed
  by black lines corresponding to
  $\langle\theta^i_t(x)\rangle\pm\sigma_{x_i}(t)$ for a) the first
  (green) and second (violet), and b) first (green) and 8th (violet)
  tagged particle in a single file with $N=3$ and $N=10$,
  respectively at three different lengths of trajectories.
  The black lines correspond to ''error bars'' on a finite-time estimate
of the probability density along a single trajectory starting in the steady-state.
c) and d): Reduced
  variance of local time of the central particle
  $\sigma^2_{x_c=1/2}(t)/\langle\theta_t^c(x_c=1/2)\rangle^2$ for
  various odd $N$ in order to preserve the symmetry. The full lines denote exact results from
  Eq.~(\ref{svar}) and dashed lines large deviation asymptotics
  Eq.~(\ref{LDsvar}). Symbols correspond to Brownian dynamics
  simulation of an ensemble of $10^6$ independent trajectories
  starting from equilibrium initial conditions.}
\label{fg2}
\end{center}
\end{figure}

The results for $\sigma^2_{x_i}(t)$ in Eq.~(\ref{svar}) for the central particle
in single files with various $N$ are
depicted in Fig.~\ref{fg2}, and reflect large fluctuations exceeding
200\% on time-scales where roughly only 50\% of the particles have collided
with their neighbors. The fluctuations display a non-trivial
dependence on $x$, which does not follow the shape of
$P_{\mathrm{eq}}(x_t)=N!x_t^{n_l}(1-x_t)^{n_r}/(n_l!n_r!)$, and reveal
striking boundary-layer effects. These deviations are clear evidence
for non-Poissonian statistics and signal that
harmonization concepts, which assume a locally equilibrated
environment \cite{harmon}, break down on the more fundamental trajectory level.
At
longer $t$, where $\sim\!$ 50-100 collisions/particle have occured, $\theta_t^i$ at different $t$ become uncorrelated according to
the central limit theorem, with $\sigma^2_{x_c}(t)$ converging to its
LD limit (\ref{LDsvar}). On these time-scales the ensemble MSD has
already saturated (compare Figs.~\ref{fg1}a and \ref{fg2}c and d).
Notably, LD asymptotics correctly capture only small
fluctuations of the order $\pm 10\%$. As noted above and confirmed
by simulations,
large deviations reflecting Gaussian statistics set an upper bound to
the fluctuations of $\theta_t^i$ (Fig.~\ref{fg2}c and d).

Single-file diffusion displays no time-scale separation in the relaxation spectrum. As a
result, the projection of dynamics onto a tagged particle coordinate
induces subdiffusion and strong non-Markovianity on time scales
$t<\lambda_1^{-1}$. The respective onset of the $\sqrt{t}$ scaling
of the tagged particle MSD shifts to shorter $t$ upon increasing
$N$ (Fig.~\ref{fg1}a). Increasing $N$ in turn leads to a high degeneracy of Bethe eigenmodes, reflecting
emerging dynamical symmetries (see Appendix F). As a result, fewer Bethe
modes are required for a convergence of the sums in
Eqs.~(\ref{svar}-\ref{scor}).   
\begin{figure}[!!ht]
\begin{center}  
\includegraphics[width=8.5cm]{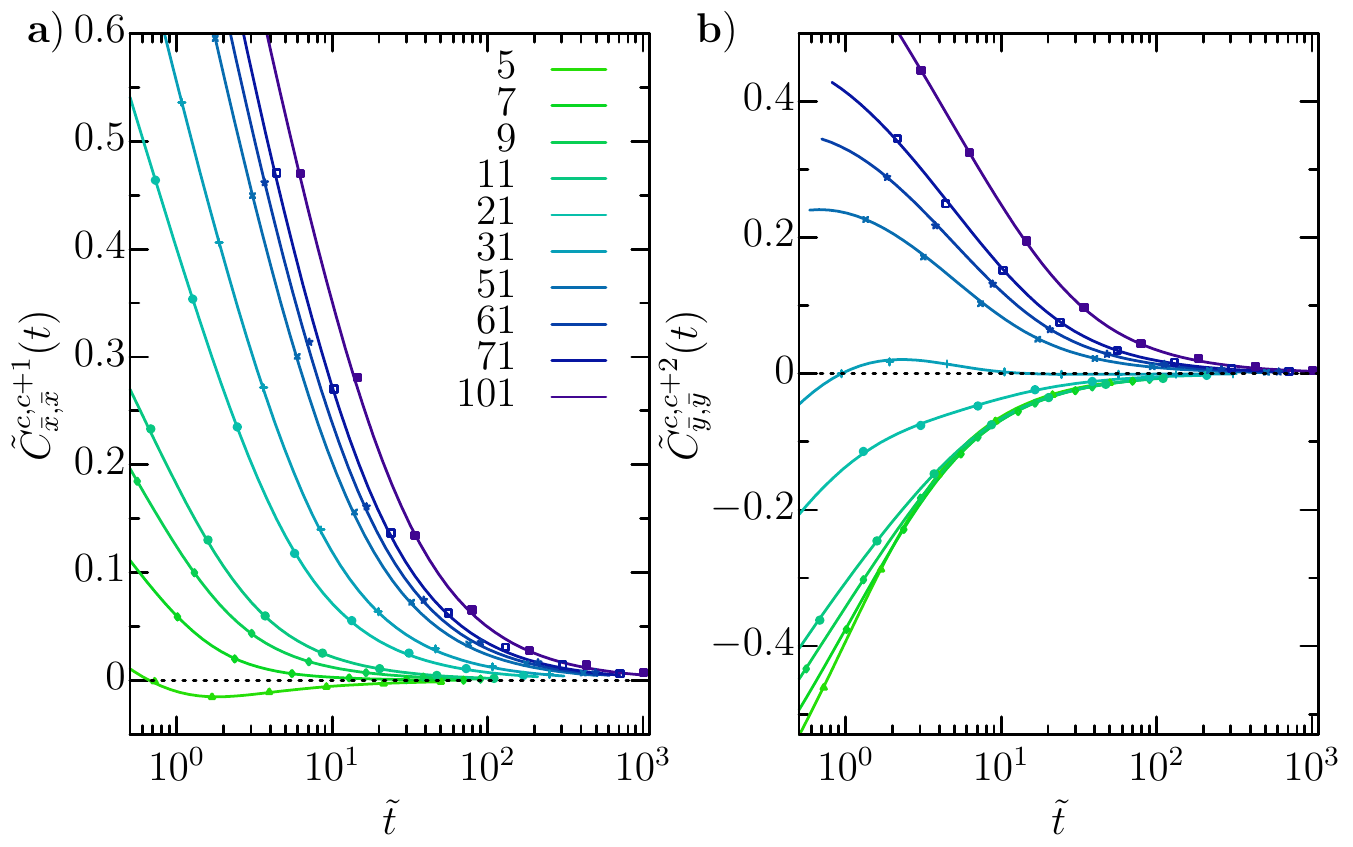}
\caption{$\tilde{\mathcal{C}}^{ij}_{xy}(t)=\mathcal{C}^{ij}_{xy}(t)/(\langle\theta^i_t(x)\rangle\langle\theta^j_t(y)\rangle)$, reduced two-tag local time correlation functions of
  the central particle $c$ and its nearest (a) and next-nearest (b)
  neighbor for different $N$. Only odd $N$ were considered to assure
  the symmetry required for a meaningful comparison. Time is expressed in
  units of the mean collision time. Lines depict the theory in
  Eq.~(\ref{scor}) whereas symbols correspond to Brownian dynamics simulations
  of $10^6$ independent trajectories
  starting from equilibrium initial conditions.}
\label{fg3}
\end{center}
\end{figure}

To gain deeper insight into the physical origin of the memory on a trajectory level we analyzed two-tag correlations between
particle histories by means of the reduced covariance of local
times
$\tilde{\mathcal{C}}^{ij}_{xy}(t)=\mathcal{C}^{ij}_{xy}(t)/(\langle\theta^i_t(x)\rangle\langle\theta^j_t(y)\rangle)$,
with $\tilde{\mathcal{C}}^{ij}_{xy}(t)\in [-1,\infty)$. Correlations
between the histories of the central particle $c$ and its nearest
(i.e. $c+1$) and
next-nearest (i.e. $c+2$) neighbors at the midpoint between the maxima of
$P_{\mathrm{eq}}(x_c)$ and  $P_{\mathrm{eq}}(x_{c+1,c+2})$ (see
Appendix E and H for details) are depicted in Fig.~\ref{fg3}.
Due to ergodicity, $\theta^i_t(x)$ become very weakly correlated at long $t$
and Gaussian statistics emerge. Consequently,
$\tilde{\mathcal{C}}^{ij}_{xy}(t)$ vanishes for long times, after
$\gtrsim 10^2$ collisions took place on average. Note that
$\mathcal{C}^{ij}_{xy}(t)$ measures correlations between
particle histories and not particle positions. The latter
never decorrelate, i.e. two-tag position correlation functions
display an algebraic decay even at equilibrium
$P^{ij}_\mathrm{eq}(x_i,x_j)=N!x_i^{n_l}(x_j-x_i)^{m_l-n_l}(1-x_j)^{m_r}/(n_l!m_r!(m_l-n_l)!)-P_{\mathrm{eq}}(x_i)P_{\mathrm{eq}}(x_j)\ne
0$, where $n_{l,r}$ and $m_{l,r}$ are the number of particles to the
left/right of the two tagged particles $i$ and $j$ (for details see
Appendix G).

Notably, we observe a transition from negatively to positively
correlated tagged particle histories upon increasing density
(Fig.~\ref{fg3}), mirroring a change in particle dynamics
from single-particle to collective fluctuations.
The driving force for
this transition can be found in an enhanced packing at higher
densities resembling a 'crystallization' transition, where invariant
tagged particle densities $P_{\mathrm{eq}}(x)$ become strongly overlapping, whereas their
respective widths shrink only very slowly (see
Fig.~\ref{eq_dens}). 
The 'critical' density, at which the behavior shifts
from negatively to positively correlated histories, depends on the
topological separation between
the two tagged particles and is shifted to higher values of $N$ for
more distant particles (compare a) and b) in Fig.~\ref{fg3}). In turn,
this reflects a growing dynamical correlation-length with increasing
$N$. As the mathematical reason for the sign-change are different
signs of leading eigenvectors entering the respective elements (see
Eq.~(\ref{element})), the transition will eventually occur of any
tagged pair. 
Moreover, upon increasing $N$, 
$\tilde{\mathcal{C}}^{ii}_{xy}(t)$ of the central particle becomes non-monotonic, with
weak anti-correlations at short $t$ turning to weak correlations
at large $t$, before reaching the LD limit of uncorrelated histories, where harmonization \cite{harmon}
ideas apply. The increasingly positive
correlations with growing $N$ reflect a persistence and a finite life-time of typical
collective fluctuations on a trajectory level, akin to glassy dynamics in
kinetically constrained models \cite{Peter_KCM}. Accordingly, positive correlations are
are not observed if we tag outer particles at external boundaries
(see Appendix H). The exact results for fluctuations and correlations of local
times in single file diffusion in Eqs.~(\ref{states}-\ref{auxBethe}),
and the explanation of the origin of broken Markovianity on a
trajectory level are our second main result.

\section{Conclusions} We established a general method
for determining exactly the variance and two-tag correlations of bounded
non-negative functionals of stationary ergodic Markov processes with a
diagonalizable propagator. The theory relates the statistics
of functionals to the relaxation eigenspectrum, and allows for
an exact treatment of non-Markovian dynamics from the corresponding
higher-dimensional Markovian embedding. It also holds for
diagonalizable irreversible dynamics, where a
broken time-reversal symmetry can cause oscillations in higher order terms in
Eqs.~(\ref{svar}-\ref{LDscor}) and/or fluctuations
exceeding the large deviation limit in Eq.~(\ref{LDsvar}).
From the spectrum of the many-body propagator obtained via the coordinate Bethe ansatz, we derived exact results for one- and two-tag local
times in single file diffusion, which unveiled non-trivial
correlations between tagged particle histories and the emergence of
collective dynamics at increasing particle densities. Going beyond
large deviation time-scales, our results revealed that harmonization concepts, assuming dynamics in-between local equilibria -- an
assumption that works well for ensemble-average observables
\cite{harmon} -- fail on the more fundamental trajectory level. This
highlights the intricate physical meaning of projection-induced memory
on the level of single trajectories, which is virtually invisible to
ensemble-average observables. Our results on local times can be readily
tested by existing particle-tracking experiments  (see e.g. \cite{traj}), and
hopefully our theory will stimulate further research directed towards
tagged particle functionals. Particularly interesting would be
extensions to tagged particle dynamics in rugged potential landscapes
\cite{David}.

\ack
We thank David Hartich for insightful discussions and critical reading of the
manuscript. The financial support from the German Research Foundation (DFG) through the
\emph{Emmy Noether Program "GO 2762/1-1"} (to AG), and an IMPRS fellowship of
the Max Planck Society (to AL) are gratefully acknowledged.

\begin{appendix}
\renewcommand{\theequation}{A\arabic{equation}}
\setcounter{equation}{0}
\section{Proof of the main result}
Let $\mathbf{x}(t)$ be an arbitrary-dimensional ergodic Markov process
on a discrete or continuous state-space. The evolution of
the probability density function evolves under the
corresponding \emph{diagonalizable} forward generator $\hat{L}$ 
(e.g. Fokker-Planck- or discrete-state master equation-type) with
invariant measure $\overline{P}(\mathbf{x})$ and the adjoint (i.e. backward) generator $\hat{L}^{\dagger}$. Let the
respective eigenspectra be $\hat{L}=\sum_{k}\lambda_k|\psi^R_k\rangle \langle\psi
^L_k|$, $\lambda_k$ and $\hat{L}^{\dagger}=\sum_{k}\lambda_k|\psi ^L_k\rangle \langle\psi
^R_k|$, $\lambda_k$ denoting the possibly degenerate and in general
complex-valued eigenvalues. Note that
$\hat{L}|\psi^R_k\rangle=\lambda_k|\psi^R_k\rangle$ and
$\hat{L}^{\dagger}|\psi^L_k\rangle=\lambda_k|\psi^L_k\rangle$, i.e. the left and right eigenstates span a bi-orthogonal eigenspace $\langle\psi^L_k|\psi
^R_l\rangle=\delta_{kl}$ \cite{Gardiner}. The forward and backward propagators of the
process can then be written as \cite{Gardiner}
\begin{eqnarray}
  P_\mathrm{f}(\mathbf{x},t|\mathbf{x}_0)&=&\langle \mathbf{x}|\mathrm{e}^{-t\hat{L}}|\mathbf{x}_0\rangle
  = \sum_{k}\langle
  \mathbf{x}|\psi_k^{R}\rangle\langle\psi^L_k|\mathbf{x}_0\rangle\mathrm{e}^{-\lambda_kt}\nonumber\\
P_\mathrm{b}(\mathbf{x},t|\mathbf{x}_0)&=&\langle \mathbf{x}_0|\mathrm{e}^{-t\hat{L}^{\dagger}}|\mathbf{x}\rangle
  = \sum_{k}\langle
  \mathbf{x}_0|\psi_k^{L}\rangle\langle\psi^R_k|\mathbf{x}\rangle\mathrm{e}^{-\lambda_kt}.  
\label{propagator}
\end{eqnarray}
Obviously, for $\hat{L}$ with a partially continuous spectrum
\footnote{The ground state is always discrete as we assume the
  existence of an invariant measure.} the sum would be replaced by the
corresponding integral,
The probability density function of a bounded functional $\varphi_t=\int_0^t\hat{V}[\mathbf{x}(\tau)]d\tau$ over all
paths starting from a (potentially non-equlibrium) steady-state and
propagating up to time $t$, is defined by the path integral
\begin{equation}
\label{PI}  
\mathcal{F}(\varphi|t)=\int \int d\mathbf{x}d\mathbf{x}_0
\overline{P}(\mathbf{x}_0)\!\!\!\!\!\int\displaylimits_{\mathbf{x}(0)=\mathbf{x}_0}^{\mathbf{x}(t)=\mathbf{x}}\!\!\!\!\!\!\mathscr{D}[\mathbf{x}(t)]\mathrm{e}^{-\mathcal{S}[\mathbf{x}(t)]}\delta(\varphi-\int_0^t\hat{V}[\mathbf{x}(\tau)]d\tau),
\end{equation}
with the corresponding stochastic action functional
$\mathcal{S}[\mathbf{x}(t)]$ of the continuous \cite{Kleinert,Satya} or
discrete state-space \cite{Frey} Markov process $\mathbf{x}(t)$, and
where we introduced the Dirac delta function $\delta(x)$.  
By means of
a straightforward vectorial generalization of the trotterization of
the the path integral (\ref{PI}) in Refs. \cite{Satya,Kac} (for the
backward and forward approach, respectively), one finds that the
generating function, corresponding to the Laplace transform
$\tilde{\mathcal{F}}(u|t)=\int_0^{\infty}d\varphi\mathrm{e}^{-u\varphi}\mathcal{F}(\varphi|t)$,
is the propagator of a tilted operator 
\begin{equation}
\tilde{\mathcal{F}}(u|t) =\langle \textendash|\mathrm{e}^{-t(\hat{L}+u\hat{V})}|\mathrm{ss}\rangle=\langle \mathrm{ss}|\mathrm{e}^{-t(\hat{L}^{\dagger}+u\hat{V})}|\textendash\rangle,
\label{tilted}
\end{equation}
where we have introduced the 'flat' $|\textendash\rangle\equiv\int
d\mathbf{x}|\mathbf{x}\rangle$ and steady states
$|\mathrm{ss}\rangle=\int d\mathbf{x}\overline{P}(\mathbf{x})|\mathbf{x}\rangle$, which are the left (right) and right (left) ground eigenstates
of $\hat{L}$ ($\hat{L}^{\dagger}$), respectively. The last equality
follows from $\tilde{\mathcal{F}}(u^{\dagger}|t)^{\dagger}=\tilde{\mathcal{F}}(u|t)$. In taking the
Laplace transform we assumed that the functional has non-negative
support (such as in the case of local times). In case the support extends to negative values one simply
needs to take the Fourier transform instead.  

The moments of
$\mathcal{F}(\varphi|t)$ at any given $t$ follow immediately from
$\langle \varphi_t^n \rangle =
(-1)^n\partial_u^n\tilde{\mathcal{F}}(u|t)|_{u=0}$, where $\langle
\cdots\rangle$ denotes the average over all trajectories starting from
a steady state and propagating up to time $t$. In case the Fourier
transform is used, a corresponding change of the prefactor is required.

For bounded functionals of ergodic Markov processes all moments are
finite, $|\langle \varphi_t^n \rangle| \le f(t)\lim_{t\to\infty}|\langle\varphi_t^n\rangle|<\infty$ with a smooth scaling function
$f(t)$, which depends on the detailed form of
$\hat{V}[\mathbf{x}(t)]$. This follows from the fact that the integral
is always over a finite time (see e.g. Eq.(\ref{local})) and hence
boundedness of the integrand assures the boundedness of the time-average observable.
Moreover, $\mathcal{F}(\varphi|t)$ obeys a
large deviation principle \cite{Varadhan,Satya2}.  
In the specific case of local times, $\hat{V}[\mathbf{x}(t)]=\mathbbm{1}^j_y[\mathbf{x}(t)]$
and $f(t)\propto t^n$ for $\langle \varphi_t^n \rangle$. The
finiteness of moments implies that $\tilde{\mathcal{F}}(u|t)$ is an
analytic (i.e. holomorphic) function of $u$ at least at and near $u=0$ for any $t$.

Note that for bounded $\hat{V}[\mathbf{x}(t)]$ we can always write
\begin{eqnarray}
\int_0^tdt'\hat{V}[\mathbf{x}(t')]=\int_0^tdt'\int
d\mathbf{x}\delta(\mathbf{x}-\mathbf{x}(t'))V(\mathbf{x})\nonumber\\=\int
d\mathbf{x}V(x)\int_0^tdt'\delta(\mathbf{x}-\mathbf{x}(t'))\equiv t\int
d\mathbf{x}V(\mathbf{x})\theta_t(\mathbf{x}).
\label{auxx}
\end{eqnarray}  

To obtain exact results for second moments we simply need to expand
$\tilde{\mathcal{F}}(u|t)$ in a Dyson series to second order in $u\hat{V}$ preserving the
time-ordering, and afterwards take the second
derivative at $u=0$. The series is guaranteed to converge, since $\hat{V}$ is bounded. Because trivially $\langle
\textendash|\mathrm{e}^{-t\hat{L}}|\mathrm{ss}\rangle=\langle \mathrm{ss}|\mathrm{e}^{-t\hat{L}^{\dagger}}|\textendash\rangle=1$, the Dyson
expansion gives \cite{Dyson} 
\begin{eqnarray}
&&\langle
\mathrm{ss}|\mathrm{e}^{-t(\hat{L}^{\dagger}+u\hat{V})}|\textendash\rangle
= 1 - u \langle\mathrm{ss}|\!\int_0^t \!\!d t'
\mathrm{e}^{-\hat{L}^{\dagger}(t-t')}\hat{V}\mathrm{e}^{-\hat{L}^{\dagger}t'}|\textendash\rangle+\nonumber\\
&&u^2 \langle\mathrm{ss}|\!\int_0^t \!\!d t' \!\!\int_0^{t'}\!\!d t'' \mathrm{e}^{-\hat{L}^{\dagger}(t-t')}\hat{V}\mathrm{e}^{-\hat{L}^{\dagger}(t'-t'')}\hat{V}\mathrm{e}^{-\hat{L}^{\dagger}t''}|\textendash\rangle\!+\!\mathcal{O}(u^3),  
\label{dyson}
\end{eqnarray}
with $t\ge t'\ge t''\ge 0$. An equivalent expansion can be obtained for
$\hat{L}$. The Dyson series (\ref{dyson}) converges for
$u\in\mathbbm{C}<\infty$.

We first prove
the convergence for any bounded \emph{linear} operator $\hat{B}$. To
this end we consider
the operator norm. Let $\Psi$  be a complete normed linear
space, and $\hat{B}: \Psi\to \Psi$. The operator norm is then defined as
$||\hat{B}||=\sup_{||\psi||=1}||\hat{B}\psi||$ with $\psi \in
\Psi$. The operator norm corresponds to the largest value $\hat{B}$
stretches an element of $\Psi$.
Since $\hat{B}$ is bounded we
have $||\hat{B}^N||\le ||\hat{B}||^N, \forall N\in\mathbb{N}$, which follows
simply from $||\hat{A}\hat{B}||\le||\hat{A}||\,||\hat{B}||$. The operator
exponential is defined as the limit
$\mathrm{e}^{\hat{B}}=\lim_{N\to\infty}\sum_{k=0}^N \hat{B}^k/k!$ and the convergence
is in operator norm, since $||\sum_{k=0}^N \hat{B}^k/k!
||\le\sum_{k=0}^N||\hat{B}||^k/k!, \forall N\in\mathbb{N}$. The series
on the right
hand side converges absolutely for any number $||\hat{B}||\in\mathbb{C}$.
Due to the
completeness of the space $\Psi$, $\mathrm{e}^{\hat{B}}$ as well belongs to a complete normed linear space, and moreover
$||\mathrm{e}^{\hat{B}}||\le\mathrm{e}^{||\hat{B}||}$. Taking
$\hat{B}=u\hat{V}$ with $u\in\mathbb{C}$ completes the proof of
convergence of the series (\ref{dyson}).

We now show that the following results also hold for \emph{bounded non-linear
functionals} $\hat{V}$ such that the
two-term Dyson expansion in Eq.~(\ref{dyson}) is always
well-behaved. Utilizing the identities in Eq.(\ref{auxx}) we find that
\begin{eqnarray}
\langle V \rangle = t\int d\mathbf{x}
V(\mathbf{x})\langle\theta_t(\mathbf{x})\rangle\\
\langle V^2 \rangle = t^2\int d\mathbf{x}\int d\mathbf{x}'
V(\mathbf{x})V(\mathbf{x}')\langle\theta_t(\mathbf{x})\theta_t(\mathbf{x}')\rangle.
\end{eqnarray}
Since both $\langle\theta_t(\mathbf{x})\rangle$ and
$\langle\theta_t(\mathbf{x})\theta_t(\mathbf{x}')\rangle$ are strictly
bounded, $\langle V\rangle$ and $\langle V^2 \rangle$ are also
bounded, because $\hat{V}[\mathbf{x}(t)]$ is by definition
bounded. For bounded $\hat{V}$ (linear or non-linear) this proves that
at least the two-term Dyson expansion is thus always finite and well
behaved (in fact all orders are a.s.).

Utilizing now the spectral expansion $\hat{L}^{\dagger}=\sum_{k}\lambda_k|\psi ^L_k\rangle \langle\psi
^R_k|$ in Eq.~(\ref{dyson}) we obtain for the first order term 
\begin{equation}
\int_0^t dt'\langle\mathrm{ss}| \sum_k |\psi_k^L\rangle\langle \psi_k^R|
\mathrm{e}^{-\lambda_k(t-t')} \hat{V} \sum_l|\psi_l^L\rangle\langle
\psi_l^R| \mathrm{e}^{-\lambda_lt'}|\textendash\rangle= t V_{00},
\label{ergmean}
\end{equation}
where we introduced $V_{lk}=\langle
\psi_l^R|\hat{V}|\psi_k^L\rangle$ and we used the fact that $\langle\mathrm{ss}|$ and
$|\textendash\rangle$ are the left and right ground states of
$\hat{L}^{\dagger}$ as well as the bi-orthogonality of the
eigenbasis. The second order term follows similarly
 \begin{equation} 
  \int_0^t\!\!\!d t'\!\!\int_0^{t'}\!\!\!d t''\!\sum_{l}V_{l0}V_{0l}\mathrm{e}^{-\lambda_l(t'-t'')}=
\frac{V_{00}^2t^2}{2}+t^2\sum_{l\neq0}\frac{V_{l0}V_{0l}}{\lambda_lt}\left(1-\frac{1-\mathrm{e}^{-\lambda_l t}}{\lambda_lt}\right).
\end{equation}
We can now trivially extend $u\hat{V}\to u\hat{A}+v\hat{B}$ for
$u,v\in\mathbb{C}$ and any two bounded operators $\hat{A}$ and
$\hat{B}$. In the specific case of tagged particle local times studied
in the main text we have $\hat{A}=\mathbbm{1}^i_y[\mathbf{x}(t)]$ and
$\hat{B}=\mathbbm{1}^j_z[\mathbf{x}(t)]$, where $\mathbbm{1}^j_y[\mathbf{x}(\tau)]=1$ if
$x_j\in dy$ centered at $y$, and zero otherwise \cite{Varadhan}.
The exact second moments are now obtained from
$\partial_{u}\partial_v\tilde{\mathcal{F}}(u|t)|_{u=v=0}$ and
$\partial^2_{u}\tilde{\mathcal{F}}(u|t)|_{u=0}$ by considering the
corresponding operators $\hat{A}$ and
$\hat{B}$.

Finally, since we consider the local time fraction
and not the total local time, we must take $t^{-2}\partial_{u}\partial_v\tilde{\mathcal{F}}(u|t)|_{u=v=0}$ and
$t^{-2}\partial^2_{u}\tilde{\mathcal{F}}(u|t)|_{u=0}$,
respectively. This completes the proof of the main general results, i.e. Eqs.~(6) and (7).

\section{Extended Phase-Space Integration in Single-File Diffusion}
The integrals involved in the evaluation of invariant measures and
matrix elements in single-file diffusion involve nesting, i.e. the
ordering of particles is strictly preserved
\begin{equation}
\mathop{\mathrlap{\int}{\,}n}_a^b f(\mathbf{x}) d\mathbf{x}=
\int_a^b dx_{1}\int_{x_1}^{b} dx_{2}\cdots\int_{x_{N-2}}^{b}
dx_{N-1}\int_{x_{N-1}}^{b} dx_{N}f(\mathbf{x})
\label{nested}
\end{equation}  
This imposes  non-trivial
topology of the phase space of the system. A tremendous
simplification is achieved through the so-called 'Extended Phase-Space
Integration' developed by Lizana and  Ambj\"ornsson, which exactly reduces the
nested high-dimensional integrals to scaled single particle integrals, e.g.
\cite{PRE:LA}:
\begin{equation}
\mathop{\mathrlap{\int}{\,}n}_a^b f(\mathbf{x})
\delta(x_{\mathrm{m}}-z)
d\mathbf{x}=\left(\prod_{i=1}^{m-1}\int_a^{z}dx_i\right)\left(\prod_{j=m+1}^{N}\int_z^{b}dx_j\right)\frac{f(x_m=z,\mathbf{x}_{N-1})}{n_l!n_r!},
\label{extended}
\end{equation}
where $n_l$ and $n_r$ are the number of particles (integrals) to the
left and right of the tagged particle $m$,
respectively. The extended phase-space integration in
Eq.~(\ref{extended}) applies to all functions $f(\mathbf{x})$, which
are invariant under the exchange $x_i\leftrightarrow
x_{i+1}$. Throughout our work all nested integrals included in the
bra-s $\langle \psi_k|$ (scalar products, matrix elements etc.) are
evaluated using the extended phase-space integration. 

\section{Eigenmode multiplicity and eigenvalue degeneracy}
As described in the main text we diagonalize the many-body Fokker-Planck
operator using the coordinate Bethe ansatz method. Each Bethe eigenstate of a Single-File of $N$ particles is uniquely defined by a tuple $k=(k_1,k_2,\dots,k_N)$. To each tuple corresponds one eigenvalue through the relation:
 \begin{equation}
  \lambda_k=\sum_{i=1}^N \pi^2 k_i^2. 
 \end{equation}
since more than one tuple may correspond to the same eigenvalue, these
are degenerate. To each tuple $k$ it is possible to associate a set
$\mathcal{K}$ containing the elements of $k$ counted once. Defining
$n_{\mathcal{K}_i}$ as the number of times the element $\mathcal{K}_i$ appears in the tuple $k$, we define the multiplicity of the eigenvectors associated to $k$ as
\begin{equation}
 m_k=\prod_i n_{\mathcal{K}_i}!.
\end{equation}

\section{Tagged particle equilibrium probability densities}
\begin{figure}[ht!]
 \begin{center}
 \includegraphics[scale=0.80]{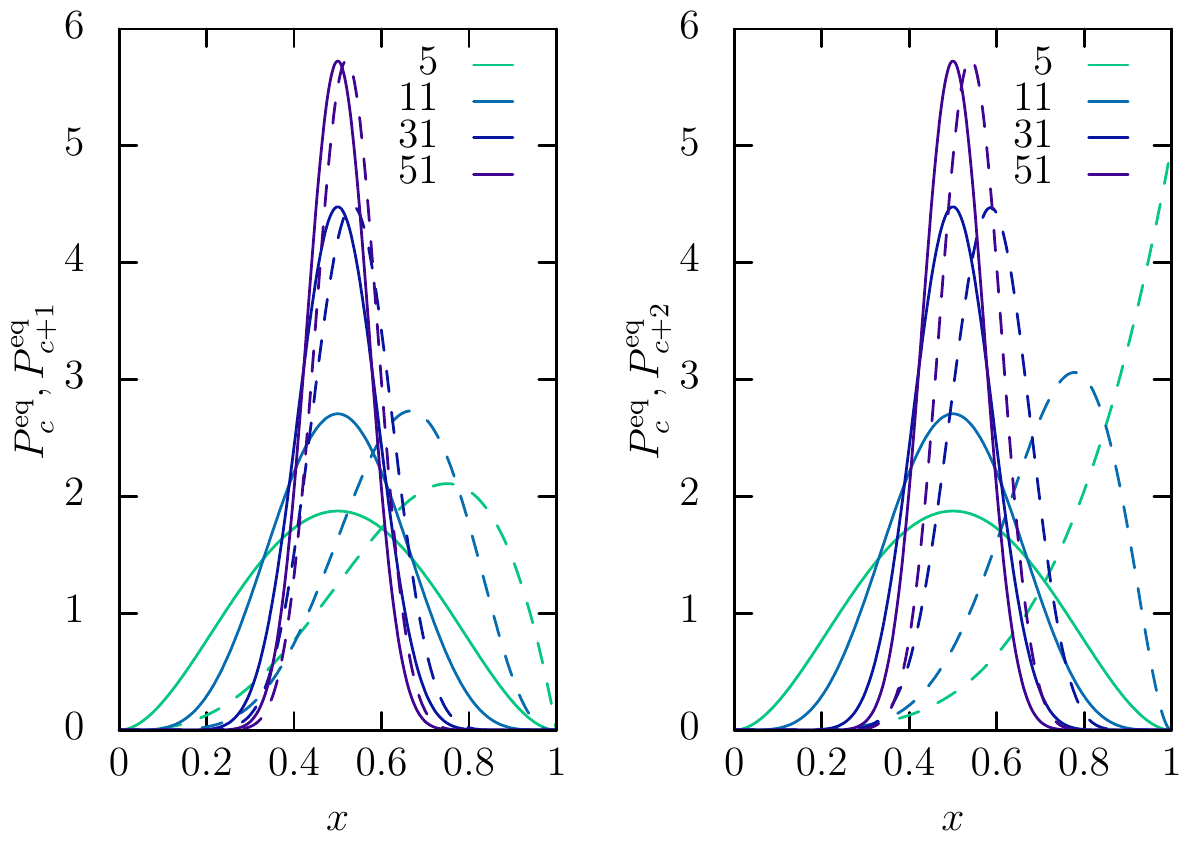}
 \caption{The solid lines represent the equilibrium probability
   density (\ref{invariant}) of the central particle and the dashed
   lines of the right nearest neighbor (left) and the next-nearest
   neighbor (right), respectively.}
 \label{eq_dens}
 \end{center}
\end{figure}

The exact tagged particle equilibrium probability density function of
the tagged particle $i$ is obtained by a nested integration of all other
particle positions
\begin{equation}
 P_i^\mathrm{eq}(x)=\mathop{\mathrlap{\int}{\,}n}_0^{\;1}
 P_{\mathrm{eq}}(\mathbf{x})\delta(x_i-x) d\mathbf{x}
 =\frac{N!}{n_l!n_r!}x^{n_l}(1-x)^{n_r},
 \label{invariant}
\end{equation}
where $n_l$ and $n_r$ are, respectively, the number of particles to
the left and to the right of the tagged particle $i$. 
Fig.~\ref{eq_dens} depicts results for $P_i^\mathrm{eq}(x)$ for the
central particle $c$ and the two nearest neighbors to the right, $c+1$
and $c+2$, respectively. The probability density of the central
particle approaches a Gaussian shape as the number density $N$
increases. For large enough $N$ the width of $P_i^\mathrm{eq}(x)$
stops decreasing appreciably, while the probability densities of
neighboring particles begin to overlap strongly. This has important
physical consequences for correlations of particle histories, as we
explained in the discussion of Fig.~3 in the main text.

\section{Reference points in the study of the density dependence of
  tagged particle local time statistics}

\begin{figure}[ht!!]
 \begin{center}
 \includegraphics[scale=0.60]{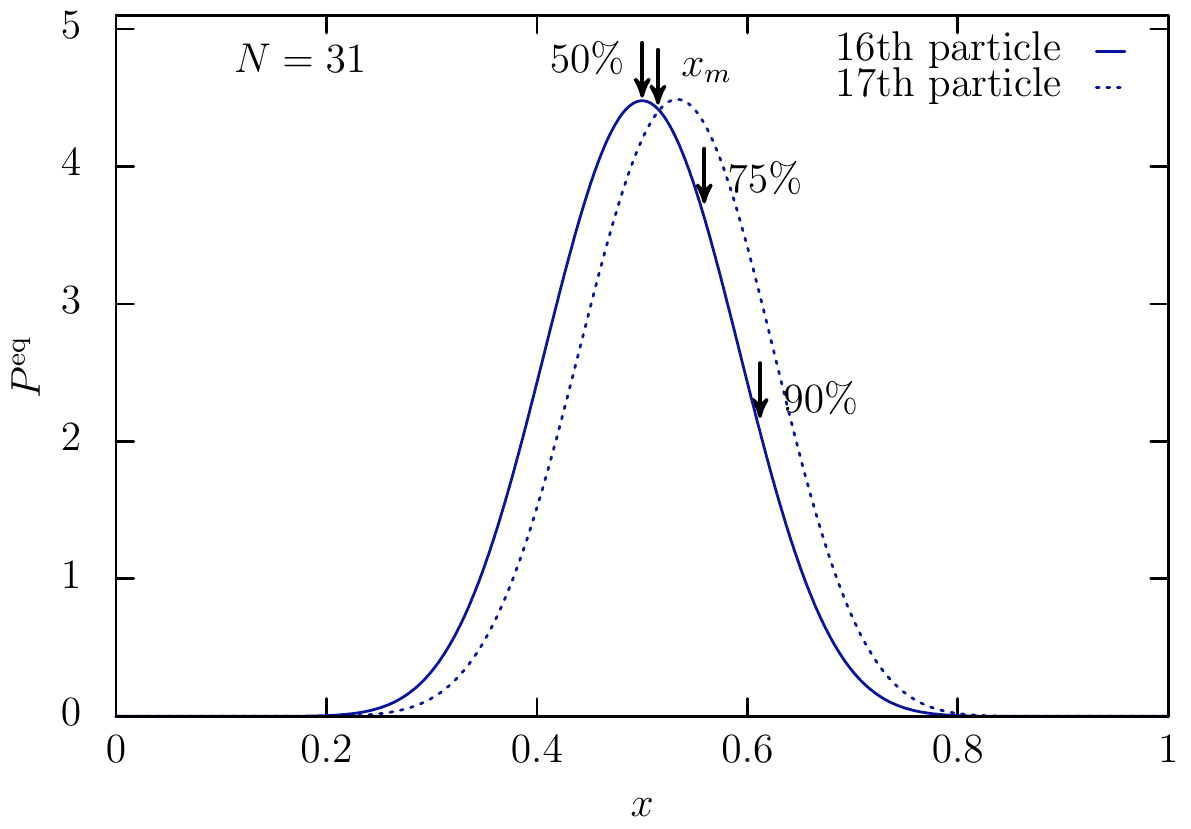}
 \caption{Invariant measures for the central particle and its nearest
   neighbor, denoting the different kinds of reference points.}
 \label{ref_dens}
 \end{center}
\end{figure}

In order to allow for a meaningful comparison of results for different
particle numbers $N$ we need to choose appropriate reference
conditions. To do so, we focus only on odd particle numbers, for which
the system is symmetric with respect to the peak of the invariant
measure of the central particle $P_{\mathrm{eq}}(x_c)$. This way a comparison
of correlations with nearest $c+1$ and next-nearest $c+2$ neighbors at different
densities is indeed consistent. Moreover, in
order to compare equilibrium and near-equilibrium tagged particle
excursions with far-from equilibrium fluctuations we choose the following
reference points with respect to $P_{\mathrm{eq}}(x_i)$: The point $x_{50}$,
in which $\int_0^{x_{50}} P_{\mathrm{eq}}(x_i)dx_i=$0.5, point
$x_{75}$, where $\int_0^{x_{75}} P_{\mathrm{eq}}(x_i)dx_i=$0.75, and
point $x_{90}$, for which  $\int_0^{x_{90}}
P_{\mathrm{eq}}(x_i)dx_i=$0.9 (see also Fig.~\ref{ref_dens}).
In the study of correlations of particle histories for two particles $i$ and $j$
we focus on the mid-point $x_m(i,j)=(x_{50,i}+x_{50,j})/2$.

\section{Convergence rates of series and eigenvalue degeneracy}
  The exact expressions for variance and covariance of local time of a tagged particles
  in Eqs.~(6) and (7) in the main text involve an infinite series, whose
  rate of convergence is difficult to predict, as it strongly depends
  on the particular position $x$ of the tagged particle under
  inspection, as well as on the number of particles $N$ and
  $\mathcal{D}(k)$, the
  degeneracy of Bethe eigenvalue $\lambda_k$. To inspect the rate of
  convergence of the series we compute the relative deviation of the
  results for the variance of local time of the central particle truncated at the $k$th Bethe eigenvalue,
  $|\sigma_{x_c}^2(t)-\sigma^2_k(t)|/\sigma^2(t)$ as a function of $k$ at different
  positions $x$ and at different lengths of trajectories $t$.   
\begin{figure}
  \begin{center}
 \includegraphics[scale=1]{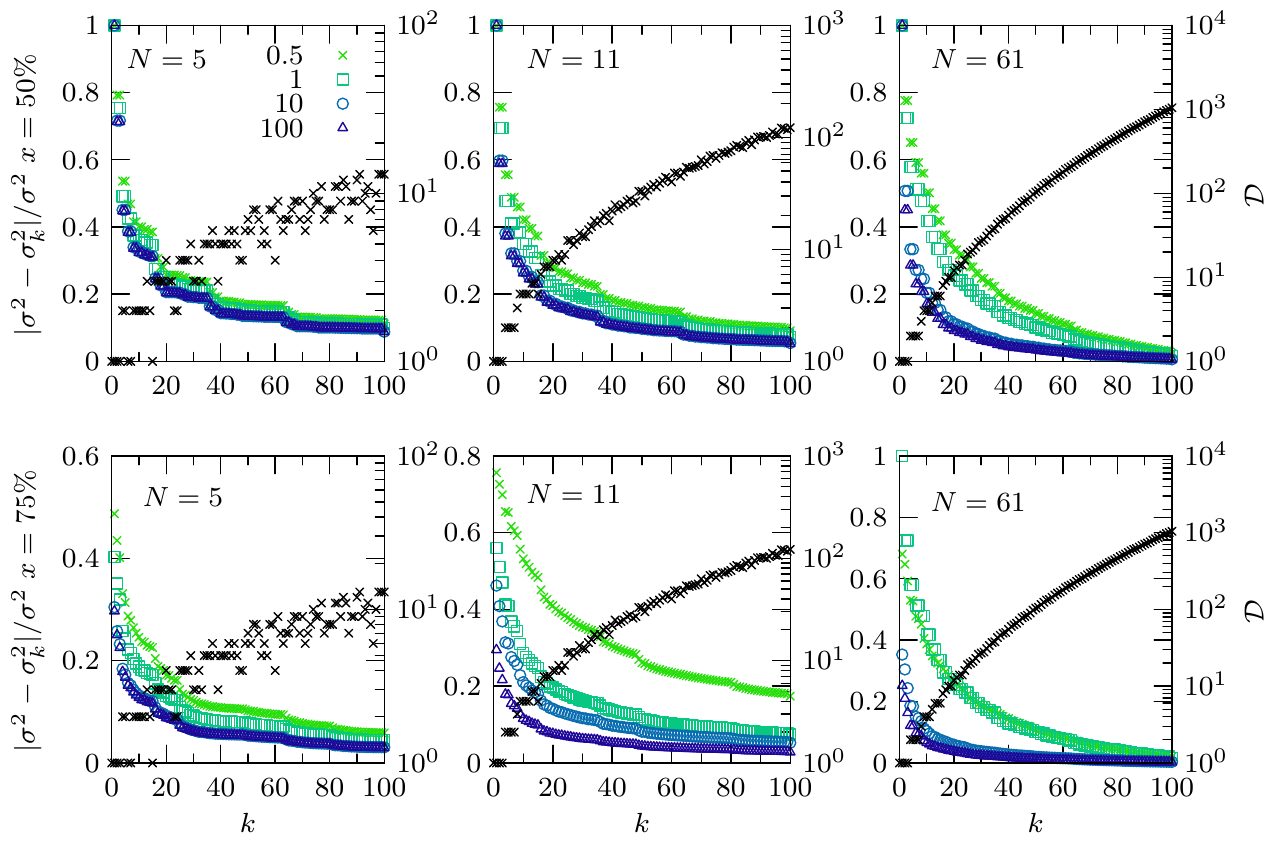}
 \caption{Analytical results for $|\sigma_{x_c}^2(t)-\sigma^2_k(t)|/\sigma^2(t)$ of the
   central particle for different particle numbers ($N$) as a function
   of $k$ at different tagging positions. The black symbols depict the
   eigenvalue degeneracy $\mathcal{D}(k)$.}
 \label{var_error}
 \end{center}
\end{figure}
Fig.~\ref{var_error} depicts how fast the series for the variance
of local time of the central particle (Eq.~(6) in the main text)
truncated at the $k$th term converges to the exact value
$k\to\infty$. n order to compare systems with different $N$ we
focused on points $x_{50}$ and $x_{75}$ of the  central particle,
with the specific values given in Tab.~\ref{position:variance}).

Intuitively, the convergence rate increases with increasing length of
the observation $t$, since faster modes must become less and less
important. The convergence rate also increases with increasing $N$,
which is due to an increasing degeneracy of lower-lying eigenvalues at larger
$N$. Degenerate low-lying eigenvalues allow
for a mixing of different collective slow modes, which become
dominant. Finally, by comparing the columns  of Fig.~\ref{var_error}
we notice that the rate of convergence also depends on the tagging
position, which in turn depends on the curvature of the modes at
different $N$.

\begin{table}
\begin{center}    
 \caption{Location of $x_{50}$ and $x_{75}$ for the central particle
   for various values of $N$.} 
 \begin{tabular}{c c c}
 \hline
  $N$	& $x_{50}$	& $x_{75}$\\
  \hline
  $5 $& $0.500$ &$ 0.641$\\
  $11$& $0.500$ &$ 0.598$\\
  $61$& $0.500$ &$ 0.544$\\
  \hline
 \end{tabular}
 \label{position:variance}
 \end{center}
\end{table}

\section{Equilibrium position correlation function}
\begin{figure}
\begin{center}
 \includegraphics[scale=0.90]{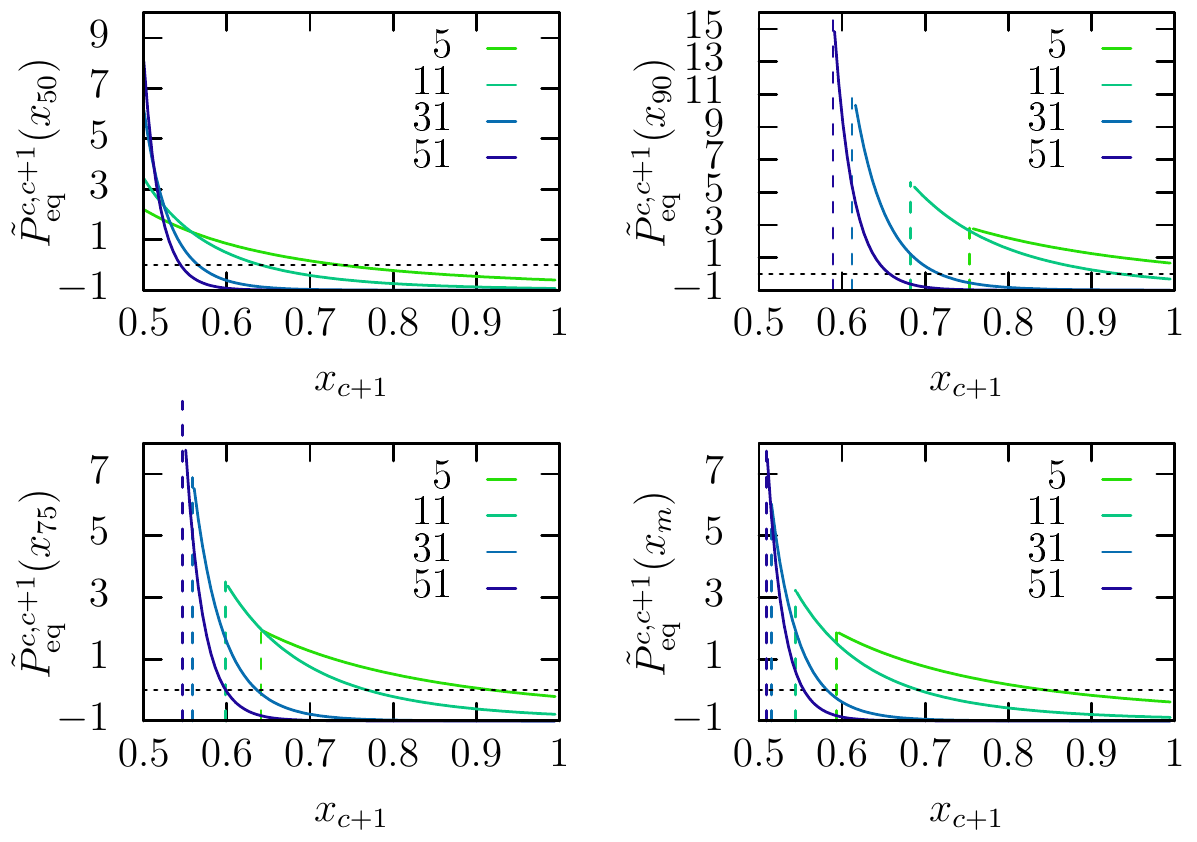}
 \includegraphics[scale=0.90]{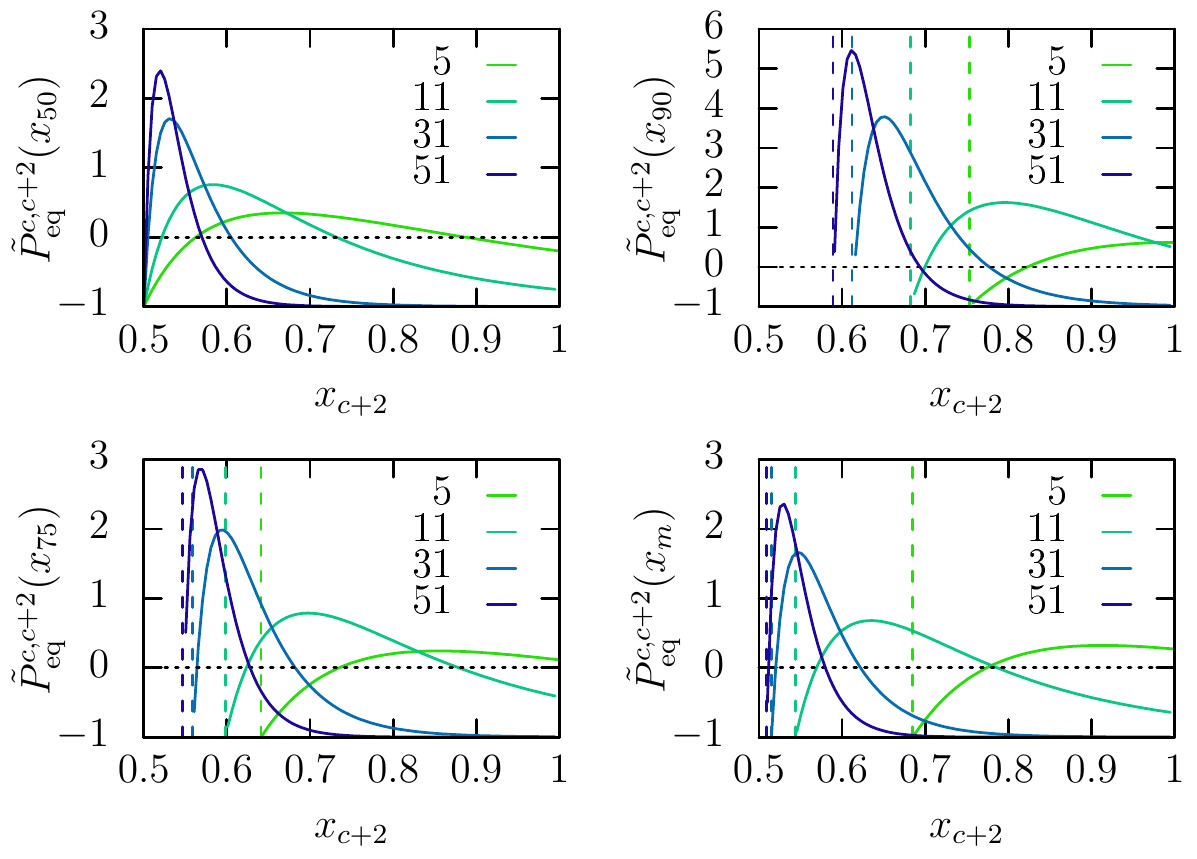}
 \caption{ $P^{ij}_\mathrm{eq}(x_i,x_j)$ between the central particle $c$
   and the right nearest and next-nearest neighbor, $c+1$ and $c+2$
   respectively, when the central particle is tagged at
   $x_c=x_{50},x_{75},x_{90}$ and when $x_c=x_m$ (see
   Tab.~\ref{position:covariance} for the exact positions). The
   vertical dashed lines are drawn at the position of the central
   particle (see Tab.~\ref{position:covariance}) to denote the
   non-crossing boundary condition.}
 \label{equilibrium:covariance}
 \end{center}
\end{figure}

In the main text we focused on the covariance of tagged particle local
times $\tilde{C}_{xy}^{ij}(t)=\frac{\tilde{C}_{xy}^{ij}(t)}{\langle
  \theta_t^i(x)\theta_t^j(t)\rangle}$ reflecting the correlations
between particle histories. We found that histories decorrelate at
long times as a consequence of the central limit theorem. Conversely,
the particle positions on the ensemble average level do not
decorrelate, not even in equilibrium. To demonstrate this we compute
exactly the pair correlation function 
\begin{eqnarray}
  P^{ij}_\mathrm{eq}(x_i,x_j)&=&\mathop{\mathrlap{\int}{\,}n}_0^{\;1}
 P_{\mathrm{eq}}(\mathbf{x})\delta(x_i-x)\delta(x_j-y) d\mathbf{x}\nonumber\\&-& \mathop{\mathrlap{\int}{\,}n}_0^{\;1}
 P_{\mathrm{eq}}(\mathbf{x})\delta(x_i-x) d\mathbf{x}\mathop{\mathrlap{\int}{\,}n}_0^{\;1}
 P_{\mathrm{eq}}(\mathbf{x})\delta(x_j-y) d\mathbf{x}\nonumber\\
&=&  \frac{N!x_i^{n_l}(x_j-x_i)^{m_l-n_l}(1-x_j)^{m_r}}{n_l!m_r!(m_l-n_l)!}\nonumber\\&-& \frac{(N!)^2x_i^{n_l}(1-x_i)^{n_r}x_j^{m_l}(1-x_j)^{m_r}}{n_l!n_r!m_l!m_r!},
\label{PCF}
\end{eqnarray}
where $n_{l,r}$ and $m_{l,r}$ are the number of particles to the
left/right of the two tagged particles $i$ and $j$. Here we want to
focus on the ensemble-average reduced pair correlation function
$P^{ij}_\mathrm{eq}(x_i,x_j)/(P_{\mathrm{eq}}(x_i)P_{\mathrm{eq}}(x_j))$
depicted in Fig.~\ref{equilibrium:covariance}. The latter is in general different from $0$, while the former goes to $0$ in the limit $t\to \infty$.
$P^{c,c+1}_\mathrm{eq}(x,y)$ depends on $N$ and for large $N$ changes
monotonically from
positive to negative correlations as a function of particle
separation. At small $N$ the correlations becomes weaker with
increasing separation, but remains positive.  Intuitively, as one particle lies in-between, the
dependence of $P^{c,c+2}_\mathrm{eq}$ on the interparticle separation
is for large $N$ non-monotonic, going from perfectly anticorrelated to correlated
and back to anti-correlation. At low $N$ $P^{c,c+2}_\mathrm{eq}$
depends  non-trivially on the interparticle separation, such that the
aforementioned terminal anticorrelation disappears for small enough
$N$. Notably, for $x_c=x_{90}$ the correlations are much stronger,
which suggest that more extensive excursions are entropically penalized
as they demand collective fluctuations.

\begin{table}
  \begin{center}  
  \caption{Reference points for the results of Fig.~\ref{equilibrium:covariance}} 
 \begin{tabular}{c c c c c}
   \hline
  $N$	& $x_{50}$	& $x_{75}$ & $x_{90}$ & $x_m$\\
  \hline
  $5 $&$0.500$&$ 0.651$ & $0.753$ & $0.593$ \\
  $11$&$0.500$&$ 0.598$ & $0.682$ & $0.544$ \\
  $31$&$0.500$&$ 0.559$ & $0.612$ & $0.515$ \\
  $51$&$0.500$&$ 0.547$ & $0.589$ & $0.509$ \\
  \hline                                           
 \end{tabular}
 \label{position:covariance}
\end{center} 
\end{table}

\section{Two-tag correlation function of local times}
In the main text (in particular in Fig.~3) we analyzed one-point two-particle histories
$\tilde{C}_{x_m,x_m}^{c,c+1,2}(t)$ at the respective mid-point positions
$x_m$ listed in Table~\ref{tab final}.

\begin{table}[ht!]
\begin{center}   
 \caption{The location of the midpoint between the central
   particle and the right nearest and second nearest
   neighbors used in Fig.~3 in the main text.} 
 \begin{tabular}{c c c}
 \hline
 $ N $	& 	$x_m$	&	$y_m$ \\
 \hline
 $5    $ &     $ 0.593 $    & $  0.685  $ \\
 $7    $ &     $ 0.567 $    & $  0.635  $ \\
 $9    $ &     $ 0.553 $    & $  0.606  $ \\
 $11   $ &     $ 0.544 $    & $  0.588  $ \\
 $21   $ &     $ 0.523 $    & $  0.546  $ \\
 $31   $ &     $ 0.515 $    & $  0.531  $ \\
 $51   $ &     $ 0.509 $    & $  0.519  $ \\
 $61   $ &     $ 0.508 $    & $  0.516  $ \\
 $71   $ &     $ 0.507 $    & $  0.514  $ \\
 $101  $ &     $ 0.505 $    & $  0.510  $ \\
 \hline
 \end{tabular}
 \label{tab final}
\end{center}  
\end{table}

To gain further insight we also compute the first and central-particle two-point
reduced correlation functions of local times $\tilde{C}_{xy}^{1,1}(t)$
and $\tilde{C}_{xy}^{c,c}(t)$
with $x$ and $y$ given in Table~\ref{position:self}. 
\begin{figure}[ht!]
 \begin{center}
  \includegraphics[scale=0.80]{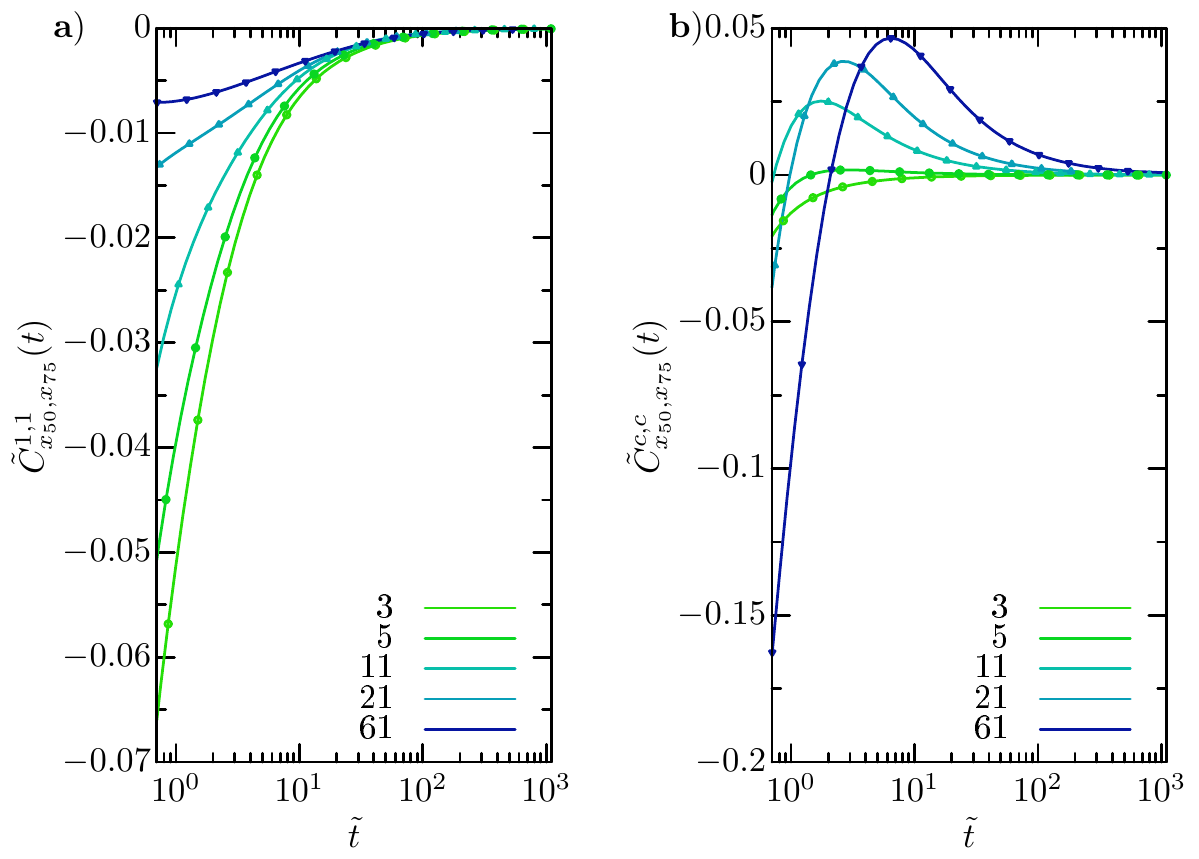}
 \end{center}
 \caption{a) $\tilde{C}_{x_{50}y_{75}}^{1,1}(t)$ of the first particle
   and b)  $\tilde{C}_{x_{50}y_{75}}^{c,c}(t)$ of the central particle.}
 \label{selfcorrelation}
\end{figure}
Fig.~\ref{selfcorrelation}). In the left plot we show the self-reduced
reveals weak anti-correlations at short times $t$ turning to weak correlations
at longer $t$, before reaching the large deviation limit of
uncorrelated histories. As already mentioned in the main text, the
correlations for the outer particles are weak and become weaker with increasing number of
particles $N$, since the outer particles are constrained between the
reflecting wall and the right nearest neighbor. In the case of the
central particle we observe further evidence of the emergence of persistent collective
fluctuations at higher densities, observed and described in the
main text. Notably here even near-equilibrium fluctuations reveal
signatures of collective behavior in the form of persistent histories
(note that we are tagging at $x_{50}$ and $x_{75}$), i.e.  $\tilde{C}_{x_{50},x_{75}}^{c,c}(t)$
turns from purely negative to weakly positive correlations.
\begin{table}
  \begin{center} 
  \caption{Reference points for the results of  Fig.~\ref{selfcorrelation}.}
 \begin{tabular}{c c c c c}
 \hline
  & \multicolumn{2}{c}{first} & \multicolumn{2}{c}{central} \\
  \hline
   $N$	& $x_{50}$	& $x_{75}$ & $x_{50}$ & $x_{75}$\\
  \hline
  $3  $&$0.206$&$0.370$&$0.500$& $0.673$ \\
  $5  $&$0.129$&$0.242$&$0.500$& $0.641$ \\
  $11 $&$0.061$&$0.118$&$0.500$& $0.598$ \\
  $21 $&$0.032$&$0.064$&$0.500$& $0.572$ \\
  $61 $&$0.011$&$0.022$&$0.500$& $0.544$ \\
  \hline
 \end{tabular}
 \label{position:self}
 \end{center} 
\end{table}
\end{appendix}

\end{document}